# GDGRU-DTA: Predicting Drug-Target Binding Affinity Based on GNN and Double GRU

Lyu Zhijian, Jiang Shaohua, Liang Yigao and Gao Min

College of Information Science and Engineering,
Hunan Normal University, Chang Sha, China

## ABSTRACT

*The work for predicting drug and target affinity(DTA) is crucial for drug development and repurposing. In this work, we propose a novel method called GDGRU-DTA to predict the binding affinity between drugs and targets, which is based on GraphDTA, but we consider that protein sequences are long sequences, so simple CNN cannot capture the context dependencies in protein sequences well. Therefore, we improve it by interpreting the protein sequences as time series and extracting their features using Gate Recurrent Unit(GRU) and Bidirectional Gate Recurrent Unit(BiGRU). For the drug, our processing method is similar to that of GraphDTA, but uses two different graph convolution methods. Subsequently, the representation of drugs and proteins are concatenated for final prediction. We evaluate the proposed model on two benchmark datasets. Our model outperforms some state-of-the-art deep learning methods, and the results demonstrate the feasibility and excellent feature capture ability of our model.*

## KEYWORDS

*Drug-Target Affinity, GRU, BiGRU, Graph Neural Network, Deep Learning.*

## 1. INTRODUCTION

So far, due to the bottleneck of technological development, the development of new drugs is more difficult, and the exploration of new uses of developed drugs has become a new hot spot. Discovering new associations between drugs and targets is critical for drug development and repurposing, However, the traditional study of drug-protein relationships in the wet laboratory [1][2] is time-consuming and expensive due to the huge range of chemical spaces to be searched, to solve this problem, some virtual screening(VS) has been proposed to accelerate the experimental drug discovery and reposition studies in silico [3], some of the more commonly used VS methods, like structure-based VS, ligand-based VS and sequence-based VS have contributed to drug development to a large extent [4][5]. However, these VS methods have their own defects in application. For example, if the structural information of the protein is unknown, the structure-based approach cannot play its role. There is still a long way to go before accurately constructing the structure of proteins, to this end, some structure-free methods have sprung up.

In recent years, with the development and maturity of deep learning technology and its great breakthroughs in the field of computer vision(CV) and natural language processing(NLP) [6][7], many people in the field of drug research have begun to turn their attention to deep learning. Moreover, with the advent of more and more biological activity data, a great deal of work based on these data has been carried out to investigate the relationship between drugs and targets. These works are usually divided into two categories, one is a binary classification-based approach, that is, to determine whether a drug and a target interact, and the other is a regression-based approach, which describes the relationship between the drug and the target by binding tightness. In binary classification-based drug-target (DT) prediction tasks, deep learning technologies seem to be used by more researches to deal with drug-target interactions (DTIs) problems. When doing DTIs prediction tasks in the past, compounds and proteins are represented using manually crafted descriptors and the final interaction prediction is made through several fully connected networks [8][9]. The problem with this approach is that the descriptors are designed from a specific perspective, that is, the design angle is too single, in addition, it remains fixed during the training process, so it cannot learn and adjust according to the results, and thus cannot extract task-related features. Therefore, some end-to-end models are proposed. Du *et al*. proposed a model called wide-and-deep to predict DTIs [10]. A generalized linear model and a deep feed-forward neural network are integrated to enhance the precise of DTIs prediction. Molecular structural information is also of great significance for feature extraction, to learn the mutual interaction features of atoms in a sequence, Shin *et al*. proposed a Transformer-based DTI model [11], which uses multi-layered bidirectional Transformer encoders [12] to learn the high-dimensional structure of a molecule from the Simplified Molecular Input Line Entry System (SMILES) string. Some researchers obtain structural information of compounds or proteins from another perspective, they represent the corresponding compounds or proteins as graphs and use graph neural networks to extract their spatial features, related work such as GraphCPI [13], Graph-CNN [14], etc.

However, the above methods have common defects, since it is a binary classification problem, the result is only yes or no, and so the distinction between compound-protein pairs is indistinguishable. In addition, many binary classification-based methods are based on setting a specific threshold as the basis of whether the drug and target interact or not. If the predictive value is higher than the threshold, it is considered interactive, otherwise it is not interactive. The deficiency of this design method is that the interaction information of many DT pairs is ignored and a proportion of these neglected information are actually significant for drug repurposing and discovery. In addition, the rationality of the threshold setting is also a factor that needs to be fully considered. Compared with the binary classification model, it seems more convincing to describe the relationship between drug and target through a regression task, the use of regression model can provide us with more information about the relationship between compounds and proteins, since continuous values can tell us how strongly the two are bound. What's more, the development of deep learning has also largely facilitated the affinity prediction of DT pairs. Related studies include KronRLS [15] and SimBoost [16], both of which based on regression and utilized the similarity information of drugs and targets to predict DTAs. DeepDTA [17] is the first framework for predicting drug and target affinity based on deep learning, which utilizes two CNN blocks to process SMILES strings of drugs and amino acid sequences of proteins, respectively. Works related to DeepDTA include WideDTA [18] and AttentionDTA [19]. The improvement of WideDTA over DeepDTA is that it combines several characters as words and proposes a word-based sequence representation method. The novelty of AttentionDTA compared to DeepDTA lies in that it proposes an attention mechanism for learning important parts of each other's sequences. In order to better capture the topological structure features of compounds, Nguyen et al. proposed GraphDTA [20] to predict drug and target affinity which utilizes RDKit technology to represent drug string sequences into graphs that could reflect its structural characteristics, and uses graph convolutional neural network to extract its spatial features. Furthermore, Lin proposed a similar approach called DeepGS [21], which uses advanced techniques to encode amino acid sequences and SMILES strings. DeepGS also combines a GAT model to capture the topological information of molecular graph and a BiGRU model to obtain the local chemical context of drug.

In this paper, we proposed a novel framework to predict DTAs. In most of the current DTA prediction research, the feature extraction of protein sequences is still dominated by CNN, this method considers the local correlation of sequences. However, most protein sequences are very long, so there are context dependences [22] in the sequence, and if we want to use CNN to capture these dependencies, then we need to use a large number of network layers. In contrast, GRU/BiGRU can capture the context dependencies of long sequences without using a large number of network layers due to its properties. Therefore, in the processing of protein representations, we interpret proteins as context-dependent time series and use GRU/BiGRU to capture the long-term dependencies of it. In the process of drug feature extraction, like GraphDTA, we still use graphs to represent drugs, and use two new graph convolution methods, namely GatedGraph and Transformer, to extract structural features of drugs. Of course, the four graph convolution methods mentioned in GraphDTA are also included for comparative experiments. Experimental results demonstrate that our model greatly improves the performance compared to previous models.

## 2. MATERIALS AND METHODS

### 2.1. Datasets

In our experimental evaluation, we used the two datasets most commonly used in DTAs prediction, namely Davis [23] and KIBA [24]. The Davis dataset contains 72 compounds and 442 proteins, along with their corresponding affinity values, where the affinity values are measured by $K_d$ values (kinase dissociation constant) and the average length of SMILES strings for compounds is 64 and that of amino acid sequences is 788. There are a total of 30056 affinity values in Davis, and they range from 5.0 to 10.8. We convert $K_d$ into the value of the corresponding logarithmic space, p$K_d$, as follows:

$$PK_{\mathrm{d}} = -\log_{10}\left(\frac{K_{\mathrm{d}}}{10^9}\right) \qquad (1)$$

The KIBA dataset contains 2116 compounds and 229 proteins, as well as 118,254 drug and target affinity values, where the affinity values range from 0.0 to 17.2. The average length of SMILES strings for compounds in KIBA is 58 and the average length of amino acid sequences is 728. The data information is summarized in Table 1.

Table 1. Summary of the benchmark datasets

| Datasets | Compound | Protein | Affinity | AC | AP | DTAsRange |
|----------|----------|---------|----------|----|----|-----------|
| Davis    | 72       | 442     | 30056    | 64 | 788 | (5.0,10.8) |
| KIBA     | 2116     | 229     | 118254   | 58 | 728 | (0.0,17.2) |

In Table 1, AC means the average length of compound strings, AP means the average length of protein amino acid sequences.

### 2.2. Overview of the proposed model

In this section, we will introduce an overview of our model. As mentioned earlier, GDGRU-DTA consists of three parts: GNN block, GRU/BiGRU block, and prediction block. After the SMILES strings of the drugs and the amino acid sequences of the proteins are given, these data are preprocessed and converted into the corresponding graph representation and feature matrix. Subsequently, the GNN block is used to extract the features of the graph representation of the drug, and the GRU/BiGRU block is used to extract the feature matrix of the protein. Finally, the extracted features of drugs and proteins are concatenated and input to the prediction block for final prediction. The overall flow of GDGRU-DTA is depicted in Fig. 1.

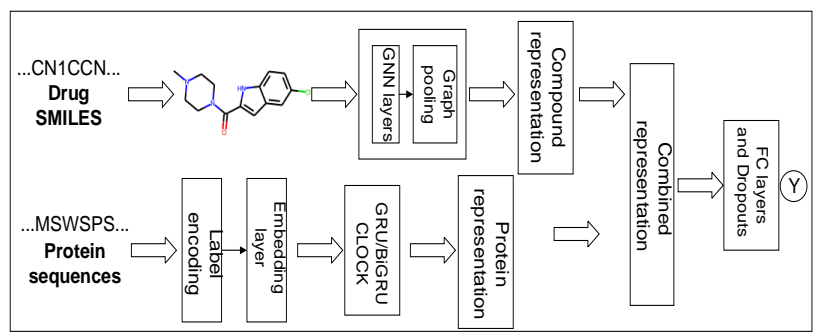

Fig. 1. Overall flow of GDGRU-DTA.

In Fig. 1, the drug and target are converted into corresponding feature representations, which are then input to the corresponding feature extraction model for feature extraction. Finally, the two extracted features are concatenated for final prediction.

### 2.2.1.  Data Preprocessing

The feature extraction of drugs and targets are two independent input channels. Before drugs and targets are input into their respective feature extraction blocks, data preprocessing is required for drugs and targets, respectively. The implementation details are as follows.

### 2.2.1.1. Drug representation

For data preprocessing of drugs, we use the same method as GraphDTA, we use the open source technology RDKit to convert the SMILES strings of drugs into corresponding 2D molecule graphs. The molecule graph is denoted as $G = (V, E)$, and the vertexes $V$ are represented as atoms and the edges $E$ are represented as bonds, where $|V| = N$ is the number of nodes in the graph and $|E| = N^e$ is the number of edges. Each atom is embedded with 78-dimensional features such as the atom's type, degree, implied valence, aromaticity, and the number of hydrogen atoms attached to the atom. The feature of the node is encoded as a one-hot vector of shape $(N, 78)$. The chemical bonds index is encoded as $(2, E)$ vector, which is used to store the edges of the undirected graph.The schematic diagram of the SMILES string of a drug converted into a two-dimensional molecule map by rdkit technology is as follows:

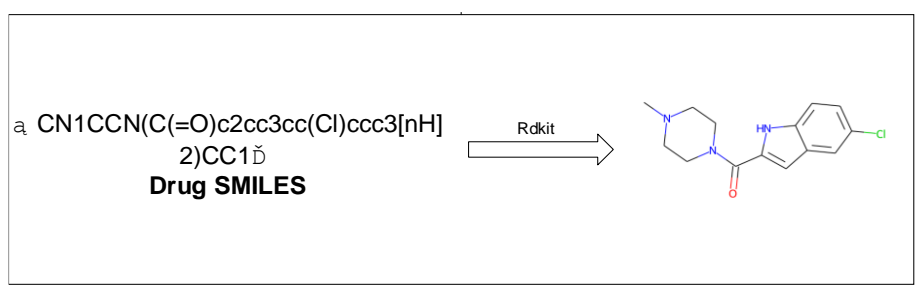

Fig. 2. Convert SMILES string to graph.

**2.2.1.2. Target representation**

The sequence length of each protein is different and varies greatly. For uniform feature representation, we fix the length of all protein sequences as 1000 according to the average length of protein sequences, if the sequence length of the protein exceeds 1000, the part more than 1000 will be cut off, and otherwise, the part less than 1000 will be padded with 0. In addition, since protein sequences are represented by different combinations of 25 amino acids, each represented by the one-letter code. We map each amino acid to an integer, and each integer is embedded as a 128-dimensional feature.

**2.2.2. GNN Blocks**

In the graph neural network block, we use two graph convolution algorithms to extract the 2D molecular graph features of drugs, namely GatedGraph and Transformer, and their details are as follows.

**2.2.2.1. GatedGraph**

GatedGraph [25] is a feature learning technique that studies graph-structured inputs, it modifies previous graph neural network work using gated recurrent units (GRU) and modern optimization techniques, and then extends to output sequences, so this method can make full use of long-distance information and fit well with our model of extracting protein features. In addition, GatedGraph has favorable inductive biases relative to purely sequence-based models when dealing with graph structure problems, and thus is a flexible and widely useful class of neural network models. The features of the node are updated as follows:

$$h_i^{(0)} = x_i \ || \ 0 \tag{2}$$

$$m_i^{(l+1)} = \sum_{j \in N(i)} e_{j,i} \cdot \Theta \cdot h_j^{(l)} \tag{3}$$

$$h_i^{(l+1)} = \text{GRU}\left(m_i^{(l+1)}, h_i^{(l)}\right) \tag{4}$$

Where in formula (2), $h_i^{(0)}$ is the input state, $x_i \in R^F$ is the feature of node i, $x_i \ // \ 0$ represents padding 0 after feature $x_i$ to the specified dimension. In formula (3), $\Theta$ is the parameter matrix to be learned, that is, the aggregation information of surrounding nodes. Formula (4) is to use a GRU unit to take the above two formulas as input and get an output, which can be functioned as a new feature of node i.

**2.2.2.2. Transformer**

Transformer [12] is a model proposed by Google researchers for seq2seq tasks, the special feature of Transformer is that it uses a lot of special layer such as self-attention in the model. Transformer breaks through the limitation that RNN models cannot be computed in parallel. Compared to CNN, Transformer does not grow with distance in the number of operations required to compute the association between two locations, and finally, self-attention can lead to more interpretable models. TransformerConv is a graph convolution method based on transformer idea [26], which takes into account the case of edge features by adopting Transformer's vanilla multi-head attention into graph learning and achieves ideal results. The feature extraction of the node is as follows:

$$x_i' = W_1 x_i + \sum_{j \in N(i)} \alpha_{i,j} \, W_2 x_j \qquad (5)$$

Where the attention coefficients $\alpha_{i,j}$ are computed via mult-head dot product attention:

$$\alpha_{i,j} = \text{softmax} \left( \frac{(W_3 x_i)^T (W_4 x_j)}{\sqrt{d}} \right) \qquad (6)$$

### 2.2.3.   GRU/BiGRU Blocks

### 2.2.3.1. GRU block

When CNN is used to extract the context dependences of long sequences, the field of view is limited due to the influence of the size of convolution kernel, and multiple CNN layers need to be used, which makes the model bloated and complex. In order to overcome the inability of CNN and RNN to deal with long-distance dependence, LSTM (Long-Short Term Memory) [27] is proposed. GRU is a very successful variant of LSTM, both of them can capture the long-term dependencies of the sequence and have comparable performance on many tasks, but GRU has a simpler internal structure and fewer parameters than LSTM, so it is more efficient when dealing with the same task, therefore, using GRU to process the time series of proteins is an obvious choice. Compared to LSTM, GRU has only two gates, namely update gate and reset gate, so it is more efficient in handling the same task. The update gate is used to control the extent to which the state information of the previous moment is brought into the current state. The larger the value of the update gate is, the more state information of the previous moment is brought into the current state. Reset gate is used to control the degree of ignoring the state information of the previous moment. The smaller the value of reset gate is, the more state information is ignored. GRU is to make a prediction in the current time step by controlling the operation of these two gates and then realizing the selection of sequence context information. The update gate $z_t$ and reset gate $r_t$ in GRU can be expressed as follows:

$$z_t = \sigma(x_t U^z + h_{t-1} W^z) \qquad (7)$$
$$r_t = \sigma(x_t U^r + h_{t-1} W^r) \qquad (8)$$

Where $\sigma$ is the sigmoid function, through which the data can be transformed into a value in the range of 0~1 to act as a gating signal. $x_t$ is the input of the current node, $h_{t-1}$ is the hidden state passed down by the previous node, and this hidden state contains the relevant information of the previous node. U and W are the corresponding weight matrices, respectively. When GRU is used to extract protein features, the feature extraction process of GRU/BiGRU blocks can be shown in Fig. 2.

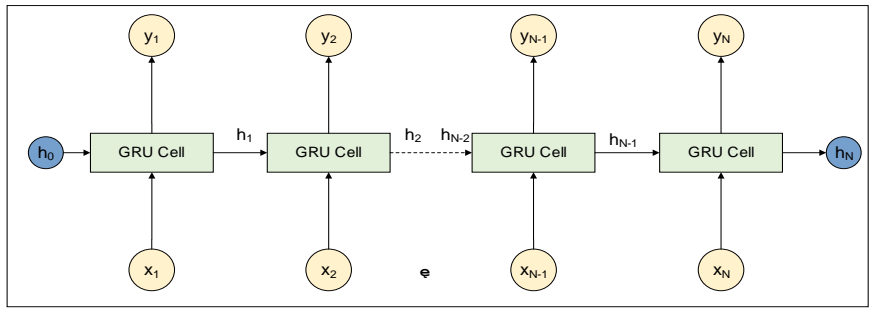

Fig. 3. GRU structure diagram.

In Fig. 2, the output of each stage is jointly determined by the hidden state of its previous stage and the input of the current stage.

### 2.2.3.2. BiGRU block

For some specific tasks, the information at a certain moment is not only related to the previous state, but also has some connection with the later state. When dealing with such problems, the traditional unidirectional GRU is obviously not competent, therefore, the bidirectional GRU is introduced. For protein sequences, we consider that the features of a certain part of the protein sequence are not only related to the previous part, but also related to the later part. Therefore, we also use a bidirectional GRU to extract the amino acid sequence features of the protein. BiGRU is composed of two unidirectional GRUs with opposite directions. At each moment, the input will fuse the outputs of the two opposite GRUs at the same time, and the output is jointly determined by these two unidirectional GRUs. The feature extraction process of BiGRU is as follows:

$$\overrightarrow{h_t} = GRU\left(x_t, \overrightarrow{h_{t-1}}\right) \qquad (9)$$

$$\overleftarrow{h_t} = GRU(x_t, \overleftarrow{h_{t-1}}) \qquad (10)$$

$$h_t = w_t\overrightarrow{h_t} + v_t\overleftarrow{h_t} + b_t \qquad (11)$$

Where the function GRU () represents converting the corresponding input to its hidden layer state. $\overrightarrow{h_t}$ and $\overleftarrow{h_t}$ represent the hidden layer state in the corresponding direction, respectively, $w_t$ and $v_t$ represent the weights corresponding to the forward hidden layer state $\overrightarrow{h_t}$ and reverse hidden layer state $\overleftarrow{h_t}$ of the bidirectional GRU at time t, respectively. $b_t$ represents the bias corresponding to the hidden layer state at time t. When BiGRU is used to extract protein features, the feature extraction process of GRU/BiGRU blocks can be shown in Fig. 3.

### 2.2.4.  Prediction block

The features of the drug and the features of the protein are concatenated after being extracted and then fed into the prediction block. The prediction block consists of two fully connected layers, each of which is followed by a Dropout of rate 0.5 to prevent over fitting. The activation function of fully connected layer is the Rectified Linear Unit (ReLU). The output of the last layer identifies the final predicted affinity value for the drug and protein.

### 2.3.  Implementation

GDGRU-DTA is implemented in Pytorch. We use the Adam optimizer with the default learning rate of 2e-4. The SMILES string for each drug is converted into 2-dimensional molecular graph where each node of the molecular graph is embedded with 78-dimensional features. GNN block consists of three stacked GNN layers with 78, 156 and 312 output features, respectively, which followed by a global max pooling layer to get the most striking features. The protein input embedding is of size 128, which means that we represent each character in amino acid sequence with a 128-dimensional dense vector. The GRU block is made up of 2 GRU layers, the first of which is followed by a Dropout of rate 0.2 and the output dimension of each layer is 8. For the BiGRU block, the number of layers of GRU is set to 1, and the output dimension is also 8. The prediction block is made up of three fully connected layers, in which the numbers of neurons are 1024, 512 and 1, respectively. The dropout rate is set to 0.5 and for the Davis dataset, the batch size is set to 128, while for the KIBA dataset the batch size is set to 512 because it is much larger than the Davis dataset, about four times larger than the Davis. Each drug and protein are converted into a 128-dimensional vector after their respective feature extraction, and are concatenated into a 256-dimensional vector for the final prediction. In this experiment, we divided the dataset into five equal parts, four of which were used as training set and one was used as test set, we deal with over fitting problem by setting up cross-validation. The number of training epochs is set to 1000. Our experiments are run on Windows 10 professional with Intel(R) Core(TM) i5-10400F CPU @ 2.90GHz and GeForce GTX 1660Ti(6GB).

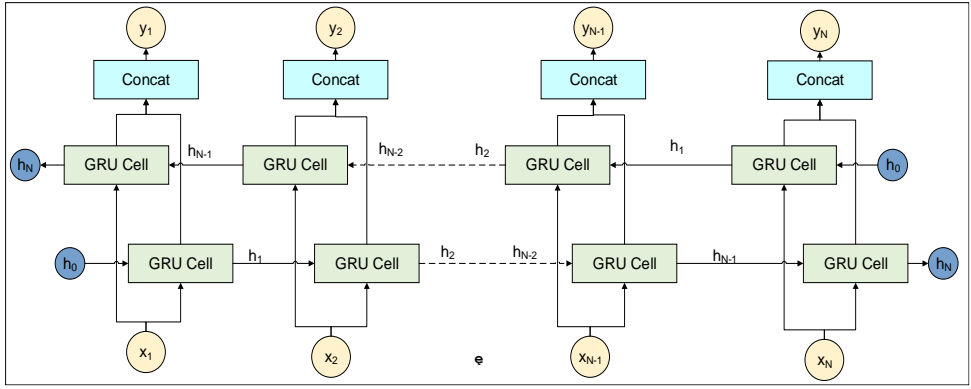

Fig. 4. BiGRU structure diagram.

In Fig. 3, the output of each stage is jointly determined by the hidden states of its previous and subsequent stages and the input of the current stage.

## 3. EXPERIMENTS AND RESULTS

### 3.1. Evaluation Metrics

MSE (Mean Squared Error), CI (Concordance Index) and $r_m^2$ (Regression toward the mean) are the most commonly used evaluation metrics in regression tasks to study drug-target interactions [15-21]. Since our research is also in this field, we continue to use these evaluation metrics, the details of each metric are as follows:

MSE is the mean square error, which is used to measure the gap between the predicted value of the model and the actual label value. The smaller the gap is, the better the performance of the model is; otherwise, the worse the performance of the model is.

$$\text{MSE} = \frac{1}{n}\sum_{i=1}^{n}(P_i - Y_i)^2 \qquad (12)$$

Where $P_i$ is the prediction value, $Y_i$ corresponds to the label value and $n$ is the total number of samples.

CI is the Concordance Index, which is a measure of whether the order of predicted binding affinity values for two random drug-target pairs is consistent with their true values, which value exceeds 0.8 indicates a strong model.

$$\text{CI} = \frac{1}{Z}\sum_{y_i > y_j} h(p_i - p_j) \qquad (13)$$

$$h(x) = \begin{cases} 1, x > 0 \\ 0.5, x = 0 \\ 0, x < 0 \end{cases} \qquad (14)$$

In (13), sample *i* has a bigger label value than sample *j*.

$R_m^2$ index is the regression toward the mean, which is used to evaluate the external predictive performance. The metric can be described as follows:

$$r_m^2 = r^2 * \left(1 - \sqrt{r^2 - r_0^2}\right) (15)$$

Where $r^2$ and $r_0^2$ are the squared correlation coefficients with and without intercept, respectively. A value of $r_m^2$ above 0.5 is considered an ideal model.

## 3.2. Results

### 3.2.1. Performance comparison of GRU/BiGRU and CNN

To demonstrate that the GRU model we use is more efficient than the CNN model in extracting protein sequence features, in this section, we compare the above two. Our experiments were carried out on the Davis database and modified on GraphDTA. We changed the way of extracting proteins from the four models of GraphDTA from CNN to GRU and BiGRU to observe their experimental results. The results are shown in Table 2, from which we can conclude that GRU/BiGRU is facilitating to capture context dependencies in sequence.

Table 2. Performances of GRU and BiGRU compared to CNN on Davis dataset

| | CI | | | MSE | | | $r_m^2$ | |
|---|---|---|---|---|---|---|---|---|
| Method | GRU | BiGRU | CNN | GRU | BiGRU | CNN | GRU | BiGRU |
| GCN | 0.899 | 0.896 | 0.880 | 0.211 | 0.220 | 0.254 | 0.712 | 0.705 |
| GAT | 0.902 | 0.903 | 0.892 | 0.218 | 0.220 | 0.232 | 0.715 | 0.706 |
| GCN-GAT | 0.895 | 0.897 | 0.881 | 0.223 | 0.232 | 0.245 | 0.697 | 0.699 |
| GIN | 0.901 | 0.896 | 0.893 | 0.214 | 0.218 | 0.229 | 0.726 | 0.714 |

As can be seen from the table, after the extraction method of protein is changed from CNN to GRU and BiGRU, both MSE, CI and $r_m^2$ are improved to varying degrees. For using the GRU model, the CI of GCN, GAT, GCN-GAT and GIN increases by 2.2%, 1.1%, 1.6%, and 0.9%, respectively, and the MSE decreases by 16.9%, 6.0%, 9.0%, and 6.6%, respectively. For using BiGRU model, the CI of GCN, GAT, GCN-GAT and GIN increases by 1.8%, 1.2%, 1.8%, and 0.3%, respectively, and the MSE decreases by 13.9%, 5.2%, 5.3%, and 4.8%, respectively. Besides, except for GCN-GAT, the $r_m^2$ values of the other methods using these two models all exceed 0.7, which indicate its excellent linear correlation and acceptability.

### 3.2.2. Comparison with other models

The GDGRU-DTA model combines GNN and RNN, and we conduct experiments on two different datasets, Davis and KIBA. The experimental results demonstrate that compared with other DTA methods, GDGRU-DTA has a huge improvement in performance. For each DTA model, we use its optimal data for comparison, the results on Davis and KIBA dataset are shown in Table 3 and Table 4 respectively.

Table 3. Results of various DTA prediction models on the Davis dataset

| Method | Protein | Compound | CI | MSE | $r_m^2$ |
|---|---|---|---|---|---|
| **Baseline models** | | | | | |
| KronRLS[15] | S-W | Pubchem | 0.871 | 0.379 | 0.407 |
| SimBoost[16] | S-W | Pubchem | 0.872 | 0.282 | 0.644 |
| WideDTA[18] | CNN | CNN | 0.886 | 0.262 | — |
| DeepDTA[17] | CNN | CNN | 0.878 | 0.261 | 0.630 |
| DeepGS[21] | GAT+Smi2Vec | CNN(Prot2Vec) | 0.882 | 0.252 | *0.686* |
| GraphDTA[20] | CNN | GNN | *0.893* | 0.229 | — |
| AttentionDTA[19] | CNN | CNN | *0.893* | *0.216* | 0.677 |
| **Proposed model – GDGRU-DTA** | | | | | |
| Transformer-BiGRU | BiGRU | GNN | **0.902** | **0.214** | **0.697** |
| GatedGraph-BiGRU | BiGRU | GNN | **0.904** | **0.214** | **0.708** |
| Transformer-GRU | GRU | GNN | **0.903** | **0.212** | **0.730** |
| GatedGraph-GRU | GRU | GNN | **0.906** | **0.207** | **0.711** |

**Table 4.** Results of various DTA prediction models on the KIBA dataset

| Method | Protein | Compound | CI | MSE | $r_m^2$ |
|---|---|---|---|---|---|
| **Baseline models** | | | | | |
| KronRLS[15] | S-W | Pubchem | 0.782 | 0.411 | 0.342 |
| SimBoost[16] | S-W | Pubchem | 0.836 | 0.222 | 0.629 |
| DeepDTA[17] | CNN | CNN | 0.863 | 0.194 | 0.673 |
| DeepGS[21] | GAT+Smi2Vec | CNN(Prot2Vec) | 0.860 | 0.193 | 0.684 |
| WideDTA[18] | CNN | CNN | 0.875 | 0.179 | — |
| AttentionDTA[19] | CNN | CNN | 0.882 | 0.155 | 0.755 |
| GraphDTA[20] | CNN | GNN | *0.891* | *0.139* | — |
| **Proposed model – GDGRU-DTA** | | | | | |
| GatedGraph-BiGRU | BiGRU | GNN | **0.892** | **0.137** | **0.775** |
| GatedGraph-GRU | GRU | GNN | **0.894** | **0.136** | **0.781** |
| Transformer-BiGRU | BiGRU | GNN | **0.894** | **0.134** | **0.780** |
| Transformer-GRU | GRU | GNN | **0.895** | **0.132** | **0.785** |

The models in the above two tables are arranged in descending order of MSE. The data for the baseline model is obtained from [15-21]. For the proposed model, two methods of drug feature extraction and two methods of protein feature extraction are randomly combined. It is not difficult to conclude from the table that the four methods of the proposed model outperform some current DTA methods to varying degrees in three indicators. In the table above, italics represent the best data of the baseline model, and bold represent the data that is better than the baseline model. In the above baseline method, KronRLS and SimBoost are traditional machine learning methods which based on similarity. DeepDTA and WideDTA are sequence-based feature extraction methods, and AttentionDTA introduces an Attention block based on it to learn mutual features. DeepGS and GraphDTA are novel in that they both use graph structure and graph convolution network to extract features.

In the analysis based on Table 3, the four approaches of the proposed model outperform the baseline model on all data, with the lowest MSE of 0.207, a 4.2% reduction compared to the lowest baseline method, and the highest CI of 0.906, compared to the highest baseline method improves by 1.5%, and the highest $r_m^2$ is 0.730, which is 6.4% higher than the highest baseline method. In addition, it can be seen from table 3 that the CI values of four methods of the proposed model all exceed 0.9, which proves that they have strong consistency, moreover, the $r_m^2$ values are all over or close to 0.7, indicating that they have strong external prediction performance. To sum up, among the above four methods of the GDGRU-DTA, the combined method of GatedGraph and GRU shows the best performance in comprehensive consideration of MSE, CI and $r_m^2$, while the combination of Transformer and BiGRU is relatively poor. The data in Table 4 shows that the performance improvement of our model on large data sets is not so obvious compared to small data sets, which indicates that our model is insufficient in some aspects, and this is a problem that we need to consider and solve.

Combining the above results of Table 3 and Table 4, we can conclude that our model has better performance than some other DTA models and has great significance for the research of DTA, and thus will greatly promote its development.

## 4. CONCLUSION

In this paper, we describe our model in detail earlier, which is an end-to-end bio-inspired deep learning-based model for DTA prediction. In this work, Since the graph structure of the drug can better represent the structural features of the drug, we represent the SMILES string of the drug as its graph structure, and use two graph convolution methods different from those used by GraphDTA, these two new graph convolution methods exhibit excellent performance on the one hand, and also demonstrate the generalization ability of the GRU/BiGRU model on the other hand. To address the feature extraction problem for long amino acid sequences, we use GRU and BiGRU to capture the long-term dependencies, in order to confirm that the model is better in protein feature extraction, we change the protein extraction method of the four models in GraphDTA to the method we used, and the results of the four models have been improved to varying degrees. GRU and BiGRU also show excellent performance when combined with our two new graph convolution methods, which demonstrate their excellent generalization ability. Finally, we combine the two newly proposed graph convolution methods and two GRU models, and compare them with the previous DTA methods and some state-of-the-art DTA methods, and the results show that our method outperforms the previous methods. Our model can greatly facilitate the affinity prediction of drugs and targets, and provide a good reference for future research.

However, there is still room for improvement in our work. For example, for the feature extraction model of drugs, our structural innovation of the model is not very large. In addition, the attention mechanism is currently widely used in the model of drug and target interaction prediction. Therefore, our next work is to investigate how to improve the structure of drug feature extraction and add attention to the proposed model to better improve its performance.